\documentclass[nofootinbib,a4paper,aps,prd,10pt,superscriptaddress,showkeys,twocolumn]{revtex4}

\usepackage{graphicx}
\usepackage{amsfonts}
\usepackage{amssymb} 
\usepackage{amsmath}
\usepackage{natbib}
\usepackage{float}
\usepackage{dcolumn}% Align table columns on decimal point
\usepackage{bm}% bold math
\usepackage{latexsym,color}
\usepackage[latin9]{inputenc}

\def\prd{Phys. Rev. D}

\def\mnras{Mon. Not. Roy. Astr. Soc.}
\def\apj{Astrophys. J.}
\def\apjl{Astrophys. J. Lett.}

\def\aap{Astron. Astrophys.}

\def\aapr{Astron. Astrophys. Rev.}

\def\araa{Annual Rev. of Astron. Astrophys.}

\def\pasj{Publications of the Astronomical Society of Japan }

\def\nar{New Astronomy Reviews}

\newcommand{\bea}{\begin{eqnarray}}
\newcommand{\eea}{\end{eqnarray}}
\newcommand{\be}{\begin{equation}}
\newcommand{\ee}{\end{equation}}

\begin{document}

\title{Accretion disk luminosity for black holes surrounded by dark matter with tangential pressure}

\author{Kuantay~\surname{Boshkayev}}
\email[]{kuantay@mail.ru}
\affiliation{National Nanotechnology  Laboratory of Open Type,  Almaty 050040, Kazakhstan.}
\affiliation{%
Al-Farabi Kazakh National University, Al-Farabi av. 71, 050040 Almaty, Kazakhstan.
}

\author{Talgar~\surname{Konysbayev}}
\email[] {talgar_777@mail.ru}
\affiliation{National Nanotechnology  Laboratory of Open Type,  Almaty 050040, Kazakhstan.}

\affiliation{%
Al-Farabi Kazakh National University, Al-Farabi av. 71, 050040 Almaty, Kazakhstan.
}

\author{Yergali~\surname{Kurmanov}}
\email[]{kurmanov.yergali@kaznu.kz}
\affiliation{National Nanotechnology  Laboratory of Open Type,  Almaty 050040, Kazakhstan.}

\affiliation{%
Al-Farabi Kazakh National University, Al-Farabi av. 71, 050040 Almaty, Kazakhstan.
}

\author{Orlando~\surname{Luongo}}
\email[]{orlando.luongo@unicam.it}

\affiliation{%
Al-Farabi Kazakh National University, Al-Farabi av. 71, 050040 Almaty, Kazakhstan.
}

\affiliation{Scuola di Scienze e Tecnologie, Universit\`a di Camerino, Via Madonna delle Carceri 9, 62032 Camerino, Italy.}

\affiliation{%
Dipartimento di Matematica, Universit\`a di Pisa, Largo B. Pontecorvo 5, Pisa, 56127, Italy.
}

\author{Daniele~\surname{Malafarina}}
\email[]{daniele.malafarina@nu.edu.kz}
\affiliation{%
Department of Physics, Nazarbayev University, Kabanbay Batyr 53, 010000 Nur-Sultan, Kazakhstan.
}

\date{\today}

\begin{abstract}
We study the motion of test particles in the gravitational field of a Schwarzschild black hole surrounded by a spherical dark matter cloud with non-zero tangential pressure and compute the luminosity of the accretion disk. The presence of non vanishing tangential pressures allows to mimic the dark matter's angular momentum while still considering a static model, which simplifies the mathematical framework. We compare the numerical results about the influence of dark matter on the luminosity of accretion disks around static supermassive black holes with the previously studied cases of isotropic and anisotropic pressures. We show that the flux and luminosity of the accretion disk in the presence of dark matter are different from the case of a Schwarzschild black hole in vacuum and highlight the impact of the presence of tangential pressures. 
\end{abstract}


\maketitle
 
%%%%%%%%%%%%%%%%%%%%%%%%%%%%%%%%%%%%%%%%%%%%%%%%%%%%%%%%%%%%%%%%%%%%%%%%%%%%%%%%%%%%%%%%%%%%%%%%%%%%%%%%%%%%%%%%%%%%%%%%

\section{Introduction}

It is widely agreed and supported by observations that the inner parts of large galaxies are characterized by the presence of supermassive compact objects \cite{2007A&A...469..405P, 2019ApJ...873...85D} as massive as millions or even billions of solar masses. The most likely candidates for such objects are supermassive black holes (BH), though extreme exotic compact objects are not excluded \cite{2019ApJ...875L...1E}. While supermassive BHs are natural options to describe the objects residing at the galactic centers their physical nature is however currently not fully-understood. 

Indeed, current models for the origin of supermassive BHs is unable to predict the observed BH distribution in terms of masses and distance \cite{Kormendy}. This poses a few cosmological issues: first, it is unclear how they have formed so massive so quickly and second, which cosmological epoch possessed the right conditions to favor their formation. 
Nevertheless, the spectroscopic properties of such supermassive BH candidates can be directly observed for several objects. In fact, we are able to study a wide number of supermassive BH candidates with masses in excess of one billion Solar masses and located in the early universe \cite{2010A&ARv..18..279V}. From such observations, we can see that they behave as \emph{accretors}, i.e., objects surrounded by accreting gas and material, forming disks and/or rotating substructures.  

The accretion disk spectra have the great advantage to be easily detectable. It is therefore natural to look at the accretion disks in order to study the properties of the central objects \cite{2019Univ....5..183D, 2002NewAR..46..239B}. 
Also, as galaxies are immersed in dark matter (DM) halos, we expect that the BH plus accretion disks must also be immersed in a DM envelope. Therefore, mathematical models of supermassive BHs surrounded by DM envelopes may be used to investigate the nature of the central object as well as the properties of proposed DM models.

To describe theoretically extreme compact objects, if we assume the validity of General Relativity\footnote{For the sake of clarity, it is not imperative to consider Einstein's gravity as many alternative theories have been proposed and their implications may be tested with similar methods as long as they fulfill the equivalence principle \cite{Capozziello:2019cav,Copeland:2006wr}.}, then the use of Einstein's field equations is essential, as one cannot avoid relativistic effects on the accretion disk luminosity. Consequently, the central object, responsible for the accretion disk's existence, produces a given geometry in its exterior. The simplest spacetime geometry is that of static black holes. 

Nevertheless, if we wish to describe an accretion disk surrounding a Schwarzschild black hole immersed in DM we may choose a non vacuum spherical solution for the exterior of the black hole. In this case, a given energy momentum tensor, whose equation of state can be  somehow specified, must be provided. In this respect, the accretion disk luminosity can significantly change due to the presence of different matter and pressure distributions outside the black hole. Hence, different choices for the matter model or its equation of state can lead to different experimental signatures \cite{Faber:2005xc}. 

In previous studies we considered DM envelopes consisting of isotropic\cite{2020MNRAS.496.1115B} and anisotropic perfect fluids\cite{2022ApJ...925..210K}. In this work we shall focus on a static model for a DM envelope composed of counter-rotating particles, the so called `Einstein cluster'~\cite{Einstein}. The model is obtained from a Schwarzschild interior solution with non vanishing tangential pressures such as the one discussed in~\cite{Florides}. Tangential pressures may be assumed to mimic the angular motion of the DM particles in the envelope. If such  motion is slow enough with respect to the time scales of the massive particles in the accretion disk the geometry of the DM envelope may be regarded as static~\cite{kumar1970non,1971GReGr...2..321B,1998PhRvD..58d1502H,2000PhRvD..61l4006J}.

In the description of the accretion disk we assume that its mass is negligible so that gas particles in the disk do not contribute to the geometry, which is uniquely determined by the BH and the DM profile. Further we assume that the DM particles have vanishing cross section of interaction with the normal baryonic matter in the disk, so that the motion of test particles in the disk may be assumed to be geodesic.

Usually DM models consider the DM particles to be in the form of dust, i.e. pressureless. In some cases, however, DM pressure may play a significant role, see, e.g., \cite{LM,Faber:2005xc,Boshkayev:2021wns} and, further, at a cosmological level it has been argued that dark energy may emerge from DM pressure, see, e.g. \cite{Boshkayev:2021uvk,Boshkayev:2019qcx,DAgostino:2022fcx,Capozziello:2017buj}.  In the model presented here, we do not consider radial pressure contributions, in the sense that we consider DM pressure small enough to be dust and the presence of tangential pressures to mimic the angular motion of the DM envelope.

To build the model we follow the standard BH accretion theory developed in Ref. \cite{novikov1973,page1974}. We then evaluate the expected observational differences between the spectrum of a BH in vacuum and that of a BH surrounded by a DM envelope. For this purpose we compute the radiative flux and spectral luminosity of the accretion disk. In addition we compare our findings with cases of non-vanishing isotropic and anisotropic pressures which were considered in earlier works~\cite{2019MNRAS.484.3325B,2020MNRAS.496.1115B}. We show that the luminosity of accretion disks surrounding BHs immersed in DM is larger than that of BHs in vacuum for a given BH mass, thus suggesting that the luminosity of observed supermassive BHs may be explained by central objects with a smaller mass that are immersed in a DM envelope. 

The paper is organized as follows. In Section \ref{sec:syscon}, we describe the system and discuss the properties of the model. In Section \ref{sec:flux_lum}, we define the radiative flux, spectral luminosity and differential luminosity, while Section \ref{sec:numres_diss} is devoted to the numerical results for the spectral properties of the accretion disks. Finally, discussions about the potential implications of our model for astrophysical BH candidates are in Section \ref{sez6}. Throughout this paper we use geometric units setting $G=c=1$.

\section{Geometry of a static black hole surrounded by a dark matter profile}\label{sec:syscon}

We consider here a general model for the inner region of a galaxy composed of a supermassive BH and a DM envelope surrounding it. In agreement with current observations, we assume DM not to interact with ordinary baryonic matter in the accretion disk, besides the gravitational interaction. To describe the physical properties of the system we consider a spherically symmetric and static metric, written in the form
\begin{equation}\label{eq:le}
d s^2=e^{N(r)} d t^2 - e^{\Lambda(r)} d r^2 - r^2 \left(d \theta^2 + \sin^2 \theta d\varphi^2\right),
\end{equation}
where,  we take $(t,r,\theta,\varphi)$ as time and spherical coordinates, respectively. The relation of the energy-momentum tensor to the metric functions is obtained from Einstein's equations and in the following we will restrict the attention to a DM cloud having only tangential pressures.

The introduction of the DM profile modifies the geometry around the black hole with respect to the vacuum case. Therefore our spacetime is made up of three parts, namely:
\begin{itemize}
\item[(i)] The galactic center is modeled by the Schwarzschild BH line element. Its mass, namely, $M_{BH}$, is a free parameter of our model and its event horizon is located at $r_g=2M_{BH}$;
\item[(ii)] The BH surrounding environment starts from an inner radius, namely $r_b>r_g$, up to the outer radius denoting the edge of the DM distribution, i.e. $r_s$. The DM mass is distributed inside this interval and its mass profile is $M_{DM}(r)$;
\item[(iii)] Finally, at $r_s$ the DM distribution reaches its maximum value and for $r>r_s$ we assume to have vacuum, i.e. we consider a Schwarzschild exterior solution with mass parameter given by $M_{BH}+M_{DM}(r_s)$, where $M_{DM}(r_s)$ is the total DM mass.   
\end{itemize}

By virtue of the above requirements, we can split the whole spacetime into three regions, 
\begin{equation}\label{eq:massprof2}
    M(r)=\left\{
                \begin{array}{lll}
                  M_{BH},  \quad \qquad \qquad \qquad r_g < r \leq r_b ,\\
                  M_{BH}+M_{DM}(r), \quad \quad r_b \leq r \leq r_s ,\\
                  M_{BH}+M_{DM}(r_s), \, \, \, \,\quad r_s \leq r .
                \end{array}
              \right.
\end{equation}
Notice that unlike the case of DM with non vanishing radial pressure, in the presence of only tangential pressure the outer boundary of the DM cloud $r_s$ is a free parameter.

The missing ingredient is clearly how DM distributes around the BH. Since we are not prescribing an equation of state we have the freedom to choose the density profile for the DM envelope. 
In the following we focus on two suitable phenomenological density distributions widely-used in the literature. The first consists in a smooth exponential sphere density profile, given by 
\begin{equation}\label{eq:den}
\rho_{Exp}(r) = \rho_0  e^{-\frac{r}{r_0}}, \quad r\in[r_b,r_s],
\end{equation}
which was discussed as a DM candidate in Ref.~\cite{2013PASJ...65..118S}. Its slope seems to correctly agree with experimental bounds inferred from the Milky Way and Andromeda galaxies. Its statistical soundness plays a key role in selecting it as plausible candidate to model DM.

The second DM density distribution considered is the Burkert profile \cite{Burkert} which is given by
\begin{equation}\label{eq:denBurk}
\rho_{Bur}(r) =\frac{\rho_{0}}{\left(1+\frac{r}{r_{0}}\right)\left(1+\left(\frac{r}{r_{0}}\right)^{2}\right)}, \quad r\in[r_b,r_s].
\end{equation}
The Burkert profile has the important advantage of not exhibiting a cusp singularity at the center $r=0$, a problem present in several popular DM profiles for galaxies such as for example the Navarro-Frenk-White profile~\cite{Navarro}. In fact, the Navarro-Frenk-White and Burkert profiles are essentially analogous, except for the very inner region. 
 
For both the density profiles considered, $\rho_0$ represents the DM density at $r=0$, whereas $r_0$ is a scale radius, determined by observations. 

Since in general for a spherical matter distribution we have that 
\begin{eqnarray}\label{eq:mprofint}
M_{DM}(r)&=&\int_{r_b}^r 4 \pi  \tilde{r}^2 \rho (\tilde{r}) \, d\tilde{r}, 
\end{eqnarray}
for the exponential sphere and Burkert profiles we obtain
\begin{eqnarray}\label{eq:mprofexden}
    M_{DM}^{Exp}&=&8 \pi r_{0}^3 \rho_{0}\bigg[e^{-x_b} \left(1+x_b+\frac{x_b^2}{2}\right)- \nonumber\\
    &&-e^{-x}\left(1+x+\frac{x^2}{2}\right)\bigg], \\
\label{eq:mprofBur}
    M_{DM}^{Bur}&=&2\pi r_{0}^{3}\rho_{0}\bigg[\frac{1}{2}\ln\left(\frac{(1+x)^{2}(1+x^{2})}{(1+x_{b})^{2}(1+x_{b}^{2})}\right)+\nonumber\\
    &&+\arctan(x_{b})-\arctan(x)\bigg],
\end{eqnarray}
with $x=r/r_0$ and $x_b=r_b/r_0$. For vanishing $r_b$ Eq.~\eqref{eq:mprofexden} reduces to the one obtained in Ref.~\cite{2013PASJ...65..118S}.

The components of the energy-momentum tensor in the absence of the radial pressures can be obtained following Ref.~\cite{Florides} as
\begin{subequations}
\begin{align}
T_{0}^{0}&=\rho(r),\quad T_{1}^{1}=0, \label{eq:enmomten}\\ T_{2}^{2}&=T_{3}^{3}=-\frac{\rho(r)M(r)}{2\left(r-2M(r)\right)}. \label{eq:enmomten2}
\end{align}
\end{subequations}

Using Einstein's equations with the line element~\eqref{eq:le}, we obtain the following expressions for the metric function inside the DM envelope

\begin{subequations}
\begin{align}
e^{-\Lambda(r)}&=1-\frac{2M(r)}{r}\,,\label{eq:Lambda}\\ 
\dfrac{dN(r)}{dr}&=\frac{2M(r)}{r(r-2M(r))}\,,\label{eq:diffN}
\end{align}
\end{subequations}
which completely determine the geometry once $M(r)$ is given. The details of the derivation of the field equations are discussed in Appendix \ref{app}.

Since the spacetime is divided into three parts we need to consider the matching at the boundaries between the different regions. The inner outer regions are vacuum solutions, while the region in between contains the DM distribution. The values of the density and metric functions at the inner boundary $r_b$ are $\rho(r_b)$ and $N(r_b)$ and $\Lambda(r_b)$ and similarly for the outer boundary $r_s$. The boundary densities can be immediately computed from Eqs. \eqref{eq:den} and \eqref{eq:denBurk} respectively as $\rho(r_b)=\rho_b$
while $N(r_b)$ is valid for both profiles and reads
\begin{equation}
N(r_b)=N_b=\ln{\left(1-\frac{r_g}{r_b}\right)+C}\,.
\end{equation}
where $C$ is a constant, unknown a priory, determined by the demand that $N(r)$ be continuous across the boundary. This requirement, namely that $g_{tt}$ is continuous at the boundary, together with the continuity of the rest of the metric tensor components, are the only necessary conditions to fulfill the Israel matching conditions \cite{1966NCimB..44....1I,1967NCimB..48..463I}.
Consequently the metric functions, $N(r)$ and $\Lambda(r)$, are evaluated as
\begin{equation}\label{eq:bc}
    e^{N (r)}=\left\{
                \begin{array}{lll}
                  \left(1-\frac{r_g}{r}\right)\tilde{C}, \, \, \, \, \, \quad\quad r_g < r \leq r_b ,\\
                  e^{N (r)}, \, \, \quad \qquad \qquad r_b \leq r \leq r_s ,\\
                   1-\frac{2 M(r_s)}{r}, \, \, \,   \, \, \, \, \qquad r_s \leq r ,
                \end{array}
              \right.
\end{equation}
where $\tilde{C}=e^{C}$ and $N (r)$ in the DM region is the  solution of Eq.~\eqref{eq:diffN} that fulfills the boundary conditions, and

\begin{equation}
    e^{\Lambda (r)}=\left\{
                \begin{array}{lll}
                  \left(1-\frac{r_g}{r}\right)^{-1},\qquad \quad r_g < r \leq r_b ,\\
                  \left(1-\frac{2 M(r)}{r}\right)^{-1}, \, \quad r_b \leq r \leq r_{s} ,\\
                  \left(1-\frac{2 M(r_s)}{r}\right)^{-1}, \, \, \, \, \, r_s \leq r .
                \end{array}
              \right.
\end{equation}

So once the mass profile is known, $\Lambda (r)$ is also known. Since the mass profile is determined unambiguously from Eq.~\eqref{eq:mprofint} and consequently Eqs.~\eqref{eq:mprofexden}-\eqref{eq:mprofBur}, defining $\Lambda(r)$, one has to numerically solve Eq.~\eqref{eq:diffN} to infer $N(r)$. 
The integration is carried out from $r_b$ to $r_s$ and the integration constant is defined from Eq.~\eqref{eq:bc} when $r=r_s$. Then $C$ is determined from the numerical solution of Eq.~\eqref{eq:diffN} when $r=r_b$. Note, since the region $r<r_b$ is not relevant for our purposes it is not necessary to explicitly evaluate $C$, which can then be absorbed into the usual Schwarzschild form by a redefinition of $t$ in the inner region. 

%%%%%%%%%%%%%%%%%%%%%%%%%%%%%%%%%%%%%%%%%%%%%%%%%%%%%%%%%%%%%%%%%%%%%%%%%%%%

\section{Radiative flux and spectral luminosity of the accretion disk}\label{sec:flux_lum}

In this section we investigate the flux and spectral luminosity of the accretion disk. 
following the framework proposed by Novikov-Thorne and Page-Thorne in Refs.~\cite{novikov1973, page1974}. 

First we focus on the angular velocity, angular momentum and energy  for particles in the disk.
Denoting $g$ as the metric determinant of the three-dimensional sub-space, composed of the set: $t,r,\varphi$, we have $\sqrt{g}=\sqrt{g_{tt}g_{rr}g_{\varphi\varphi}}$. So, one can write  
\begin{eqnarray}
\Omega(r)&=& \frac{d \varphi}{dt}=\sqrt{-\frac{\partial_{r} g_{tt}}{\partial_r g_{\varphi\varphi}}}, \label{eq:angvel}\\
L(r)&=&-u_{\varphi}=-u^{\varphi}g_{\varphi\varphi} =-\Omega u^{t}g_{\varphi\varphi}, \label{eq:angmom}
\end{eqnarray}
and
\begin{eqnarray}
E(r)&=&u_{t}=u^{t}g_{tt}, \label{eq:energy} \\
u^{t}(r)&=&\dot{t}=\frac{1}{\sqrt{g_{tt} + \Omega^{2} g_{\varphi\varphi}}},
\end{eqnarray}
where $\Omega=\Omega(r)$ is the orbital angular velocity,  $L=L(r)$ the orbital angular momentum per unit mass, $E=E(r)$ the energy per unit mass of the test particle and $u^\mu$ is the particle's four-velocity.

To implement the above in the numerical setup, it is helpful to define dimensionless counterparts to the above quantities. Thus, denoting with the superscript ${}^*$ the dimensionless quantities, we define 
\begin{subequations}
\begin{align}
\Omega^*(r)&=M_T\Omega(r)\,,\\
L^*(r)&=\frac{L(r)}{M_T}\,,\\ 
E^*(r)&=E(r)\,,\\  \mathcal{F}^*(r)&=M_T^2\mathcal{F}(r)\,,
\end{align}
\end{subequations}
where $M_T$ is the total mass of the system which corresponds to $M_T=M_{BH}+M(r_s)$ for the case of a black hole and DM distribution and to $M_T=M_{BH}$ for the case of a black hole in vacuum, i.e. without DM.

The radiative flux, $\mathcal{F}$ emitted from the disk is defined as
\begin{equation}
 \label{eq:flux}
\mathcal{F}(r)=-\frac{\dot{{\rm m}}}{4\pi \sqrt{g}} \frac{\Omega_{,r}}{\left(E-\Omega L\right)^2 }\int^r_{r_{ISCO}} \left(E-\Omega L\right) L_{,\tilde{r}}d\tilde{r}\,,
\end{equation}
which clearly depends on the quantity $\dot{{\rm m}}$, namely the \emph{disk's mass accretion rate} which, while being an essential ingredient in the computation, is also a rather arbitrary quantity. The simplest approach is to take it as constant. 
Notice that $\mathcal{F}$ is not an observable quantity and therefore we must use it to build some quantity that can be measured.
Once $\dot{{\rm m}}$ has been fixed we can construct another important quantity, namely  the differential luminosity at infinity $\mathcal{L}_{\infty}$, which is physically interpreted as the energy per unit time reaching an observer at infinity. 
Following \cite{novikov1973, page1974} we have
\begin{equation}
 \label{eq:difflum}
\frac{d\mathcal{L}_{\infty}}{d\ln{r}}=4\pi r \sqrt{g}E \mathcal{F}(r).\\
\end{equation}

Now in order to obtain a spectral luminosity, as a function of radiation frequency $\nu$, we need to choose an emission profile for the disk.
The simplest approach consists in approximating the emission with that of a black body. 
We can then convert the differential luminosity as a function of $r$ into the spectral luminosity which depends on the frequency $\nu$ via \cite{2020MNRAS.496.1115B}
\begin{equation}
 \label{eq:speclum}
\nu \mathcal{L}_{\nu,\infty}=\frac{60}{\pi^3}\int^{\infty}_{r_{ISCO}}\frac{\sqrt{g }E}{M_T^2}\frac{(u^t y)^4}{\exp\left[u^t y/\mathcal{F}^{*1/4}\right]-1}dr,
\end{equation}
where we set the dimensionless variable $y=h\nu/kT_*$, with $h$ the Planck constant, $k$ the Boltzmann constant and $T_*$ the characteristic temperature defined from the Stefan-Boltzmann law, namely 
\begin{equation}
\sigma T_*=\frac{\dot{{\rm m}}}{4\pi M_T^2}\,,
\end{equation}
with $\sigma$ being the Stefan-Boltzmann constant.
\begin{figure*}[ht]
\begin{minipage}{0.49\linewidth}
\center{\includegraphics[width=0.97\linewidth]{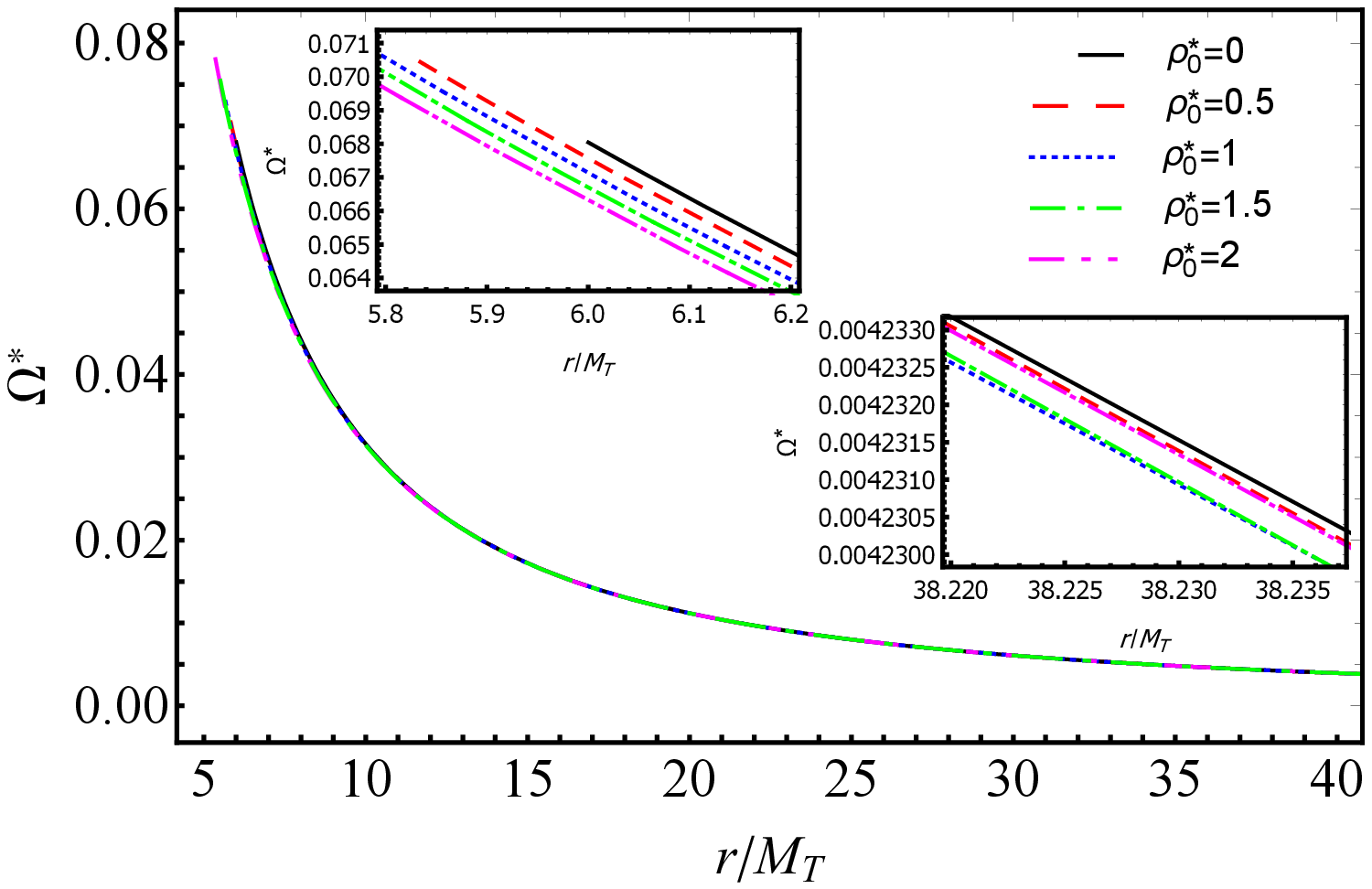}\\ }
\end{minipage}
\hfill 
\begin{minipage}{0.49\linewidth}
\center{\includegraphics[width=0.97\linewidth]{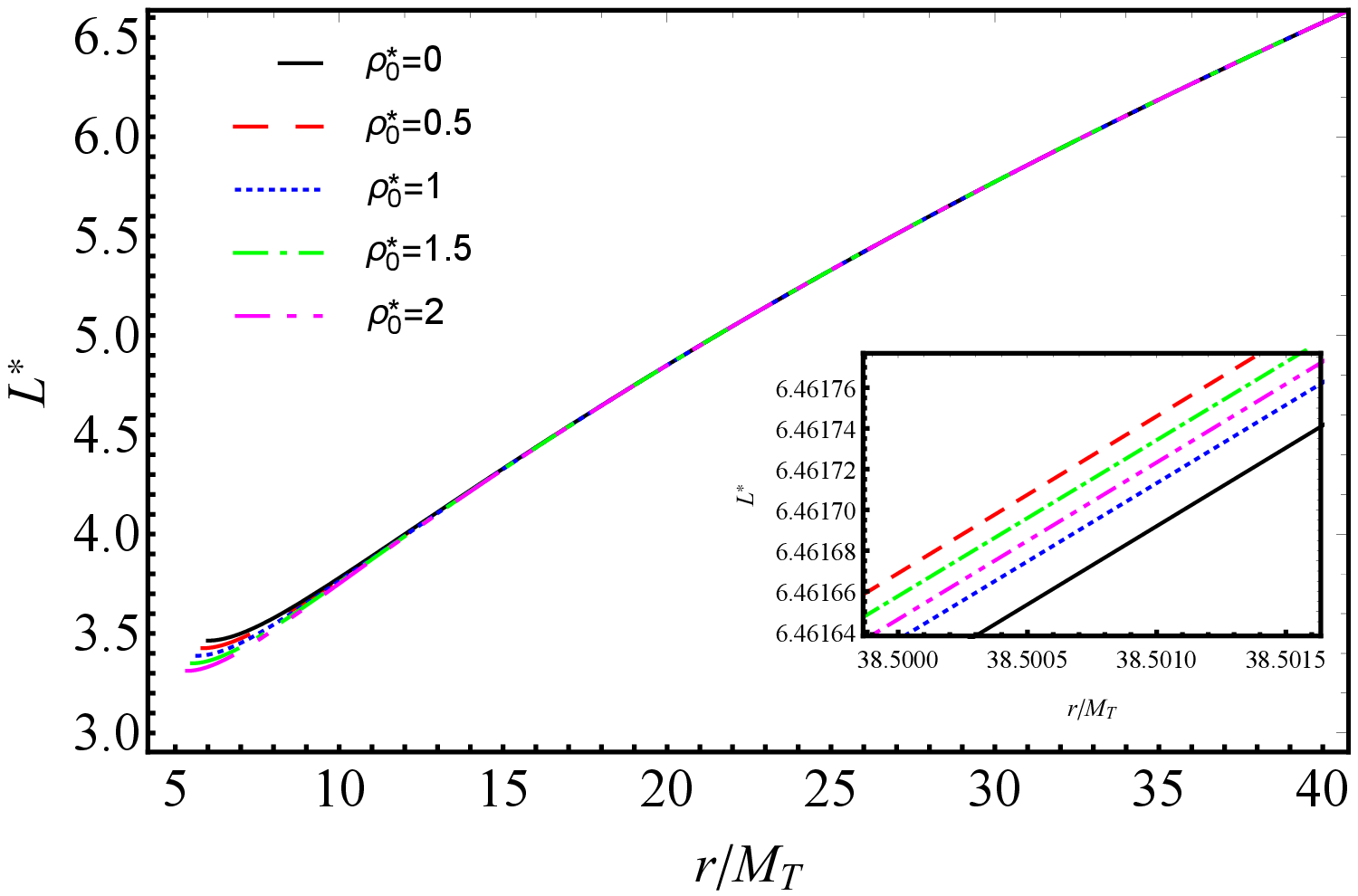}\\ }
\end{minipage}
\caption{Numerical evaluation of orbital parameters of test particles in the accretion disk in the presence of the exponential sphere DM profile as functions of $r/M_T$. Left panel: orbital angular velocity $\Omega^*$. Right panel: orbital angular momentum $L^*$. In both figures the solid curves represents the case of a static black hole without DM and dashed curves show the presence of DM with different values of $\rho_0^*=\rho_0/(10^{-5} AU^{-2})$. The corresponding plots for the Burkert density profile can also be obtained and do not show qualitative differences. }
\label{fig:omega_L}
\end{figure*}
\begin{figure*}[ht]
\begin{minipage}{0.49\linewidth}
\center{\includegraphics[width=0.97\linewidth]{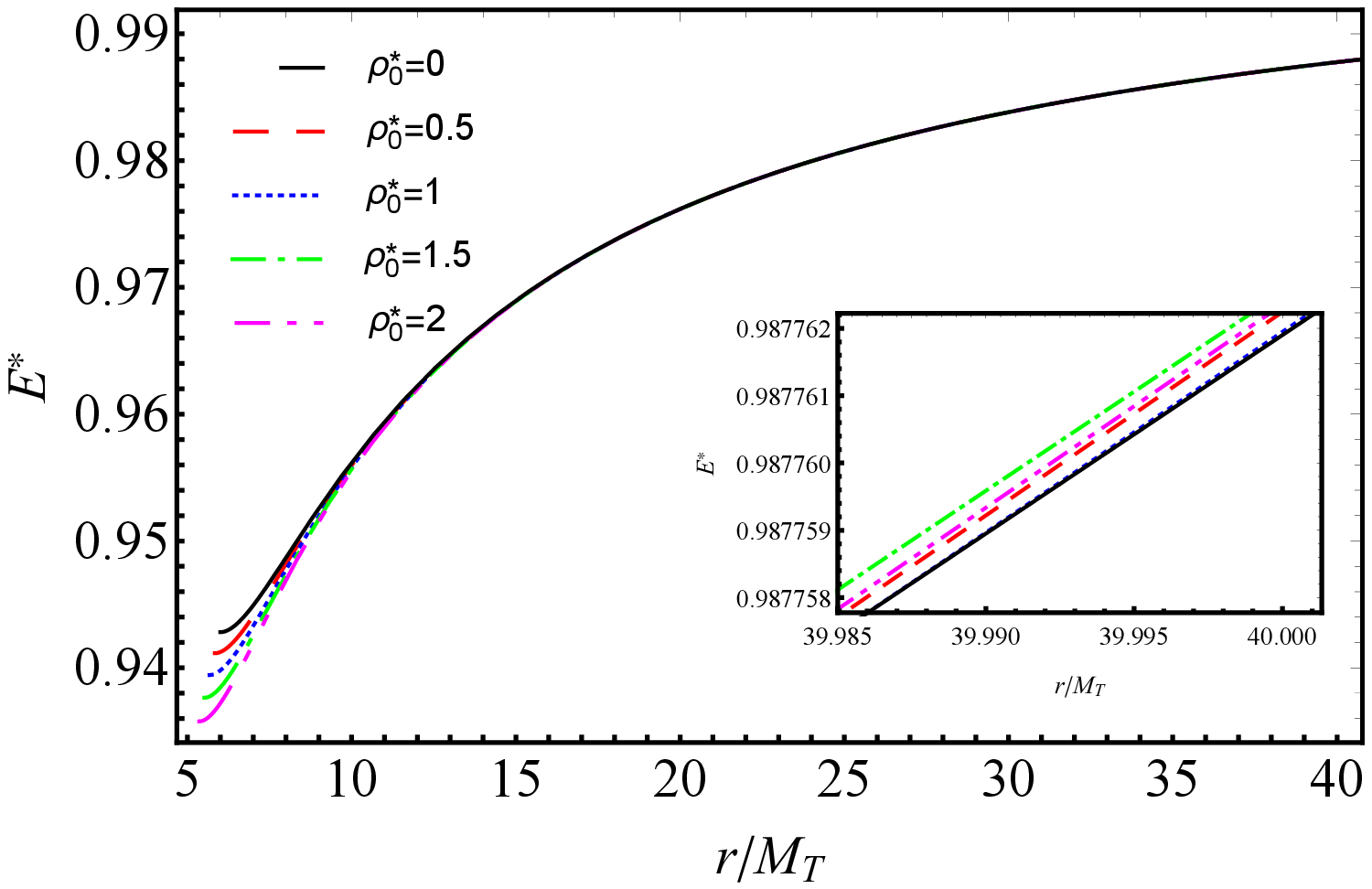}\\ } 
\end{minipage}
\hfill 
\begin{minipage}{0.49\linewidth}
\center{\includegraphics[width=0.97\linewidth]{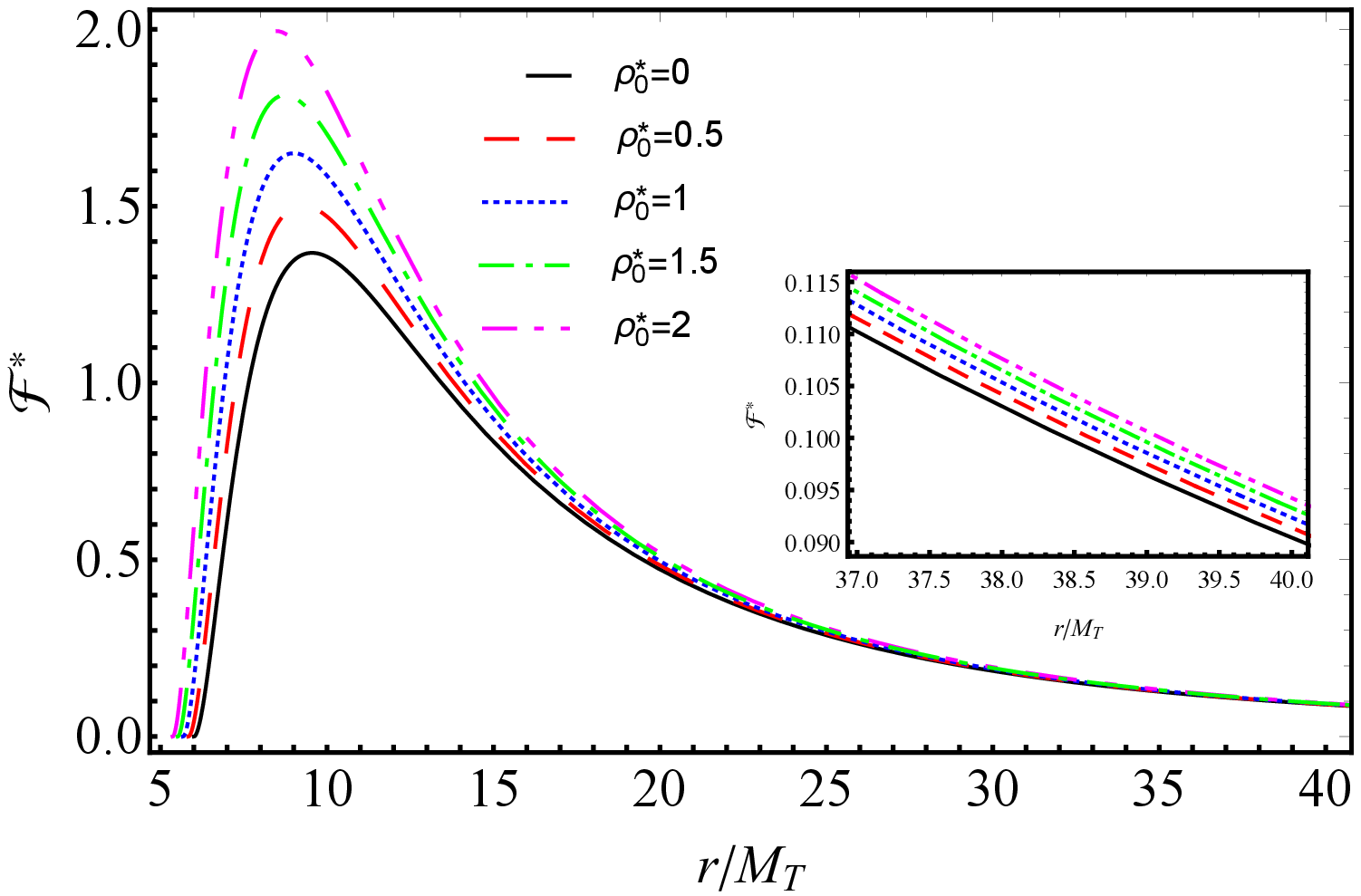}\\ }
\end{minipage}
\caption{Left panel: Energy of test particles $E^*$ in the accretion disk of a BH surrounded by DM as a function of dimensionless radial coordinate $r/M_T$. Right panel: Radiative flux $\mathcal{F^*}$, divided by $10^{-5}$, emitted by the accretion disk as a function of $r/M_T$. In both figures the solid curves represents the case of a static black hole without DM and dashed curves show the presence of DM with different $\rho_0^*=\rho_0/(10^{-5} AU^{-2})$. The plots correspond to the exponential sphere density profile. The corresponding plots for the Burkert density profile can also be obtained and do not show qualitative differences.}
\label{fig:E_F}
\end{figure*}

\section{Numerical results}\label{sec:numres_diss}

To get numerical results we need to set the free parameters of the model. Having in mind the qualitative description of the environment surrounding supermassive black holes we have set 
\begin{subequations}
\begin{align}
&M_{BH}=5\times 10^8 M_\odot\approx4.933 AU\,,\\
&r_b=\frac{11M_{BH}}{2}\approx27.133 AU\,,\\
&r_0=10{\rm AU}\,, \\
&\rho_0^*=\rho_0\times (10^{5}\rm{AU^{2}})\,,
\end{align}
\end{subequations}
where $M_\odot$ is the Sun's mass, $r_b$ is the inner boundary where the DM distribution starts and $\rho_0$ remains a free parameter that we can vary. 
In the following the value of the DM density parameter $\rho_0^*$ is 
chosen to be $\{0.5, 1, 1.5, 2\}$. 

The metric functions in the DM envelope are obtained by numerically solving Eq.~\eqref{eq:diffN} with the appropriate boundary conditions. 
Then we calculate the orbital parameters of test particles in the accretion disk and the spectral properties of the disk's emission and compare the results with the case of a black hole in vacuum and a black hole surrounded by other kinds of DM profiles. 

\begin{figure*}[ht]
\begin{minipage}{0.49\linewidth}
\center{\includegraphics[width=0.97\linewidth]{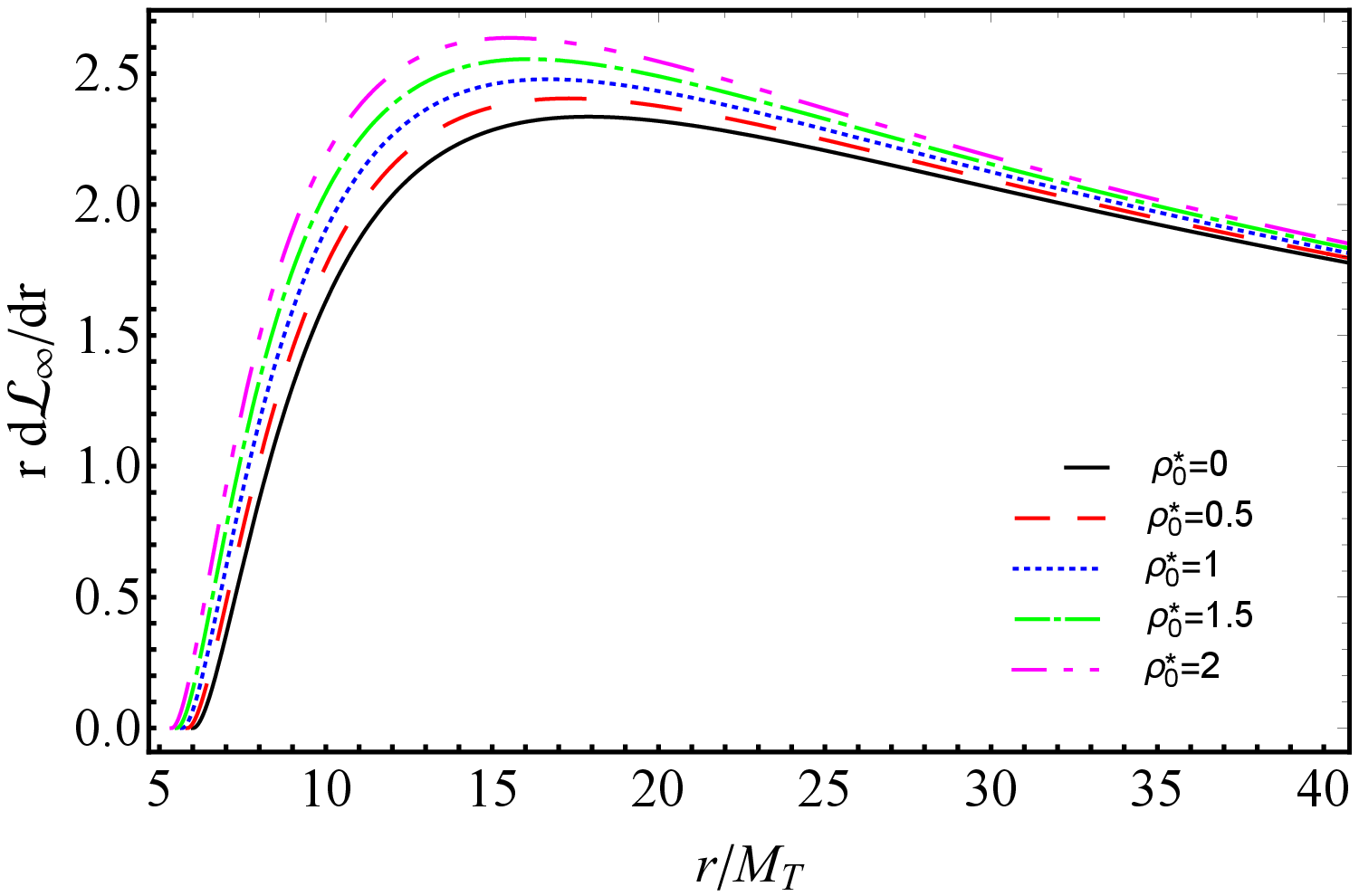}\\ } 
\end{minipage}
\hfill 
\begin{minipage}{0.50\linewidth}
\center{\includegraphics[width=0.97\linewidth]{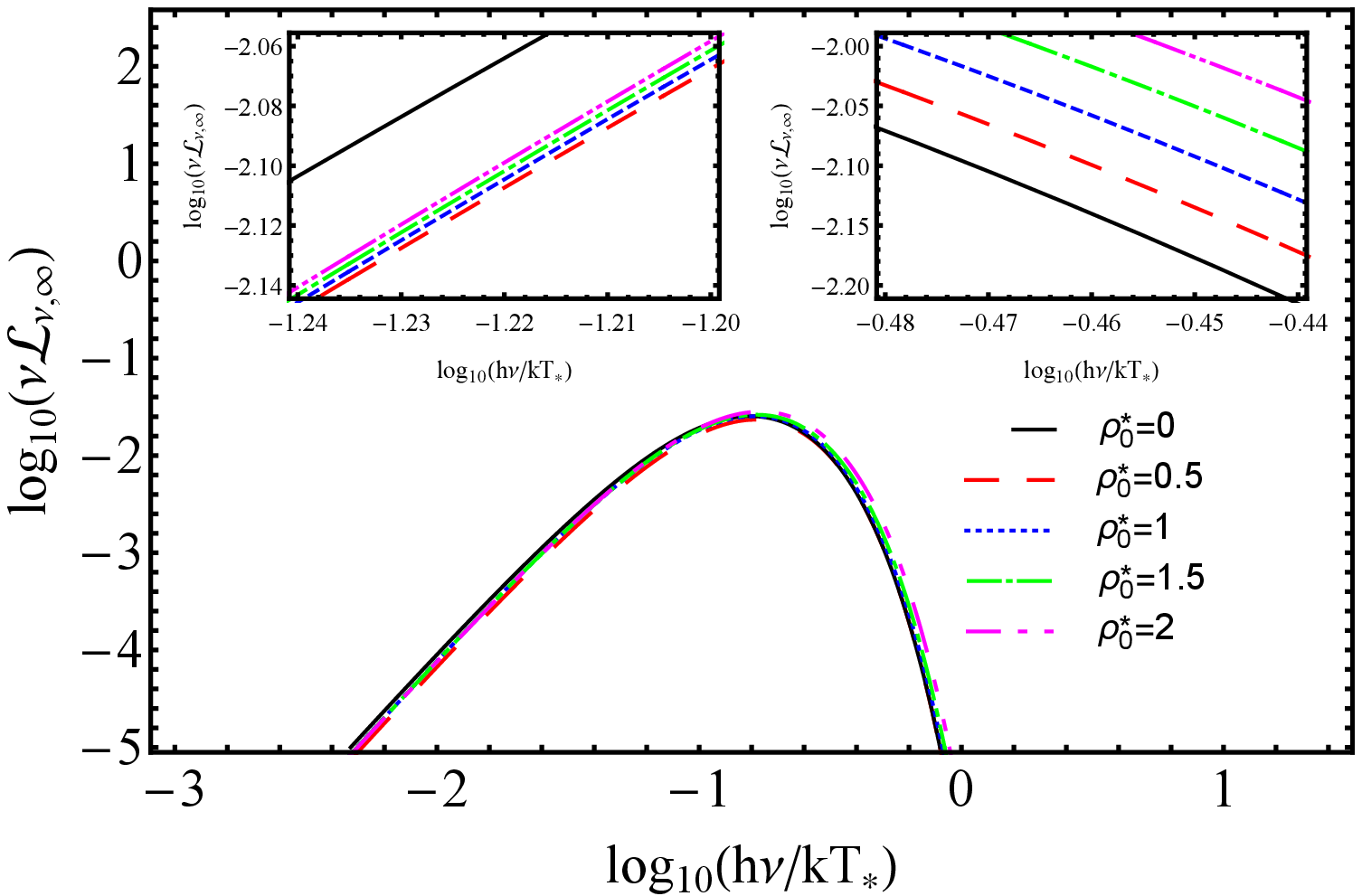}\\ }
\end{minipage}
\caption{Left panel: Numerical evaluation of the differential luminosity scaled in powers of $10^{-2}$ as a function of $r/M_T$. Right panel: Numerical evaluation of the spectral luminosity of the accretion disk as a function of $h\nu/kT_*$ , i.e. as a function of frequency. In both figures the solid curves represents the case of a static black hole without DM and dashed curves show the presence of DM with different $\rho_0^*=\rho_0/(10^{-5} AU^{-2})$. Here we used the exponential density profile only.}
\label{fig:diff_specLum}
\end{figure*}

In Fig.~\ref{fig:omega_L} we plot angular velocity and angular momentum (left and right panels respectively) of test particles in the accretion disk. In the figure the DM density profile is the exponential sphere given by Eq.~\eqref{eq:den} and, for comparison, we show the corresponding curves for test particles around a BH in vacuum, given by $\rho_0^*=0$. Noticeably, the curves of the angular velocity in the presence of DM lie below the pure vacuum case at all range of radius while the curves of orbital angular momentum lie below the vacuum case at small radii and above at large radii.

The angular momentum is important since, by solving the equation $dL/dr=0$, one obtains the smallest radius where stable circular orbits are allowed to exist, namely $r_{ISCO}$ which determines the inner edge of the accretion disk. The inner edge of the disk in turn appears in the integral of Eq.~\eqref{eq:flux}, and therefore is related to observable quantities. 

In Fig.~\ref{fig:E_F} we plot dimensionless energy of test particles in the accretion disk (left panel) and dimensionless radiative flux (right panel) given in Eq.~\eqref{eq:flux} emitted by the accretion disk as a function of the dimensionless radial coordinate $r/M_T$. For small values of $r/M_T$ the energy of test particles is smaller in the presence of DM with respect to a black hole in vacuum. But for larger $r/M_T$ the energy is larger than in vacuum. On the other hand one can observe that the dimensionless radiative flux in the presence of DM is always larger than the pure vacuum case for any $r/M_T$.

In Fig.~\ref{fig:diff_specLum} we plot the differential luminosity (left panel) as a function of $r/M_T$ and spectral luminosity (right panel) of the accretion disk as a function of $h\nu/kT_*$. Again, for all values of $r/M_T$ the differential luminosity is larger in the presence of DM with respect to the vacuum case. Though all the quantities i.e. $\Omega$, $L$, $E$, $\mathcal{F}$ and $r d\mathcal{L}_{\infty}/dr$ are crucial and present theoretical interest, the quantity that may be measured through observation is the spectral luminosity. 

In this regard, we show in Fig.~\ref{fig:diff_specLum} (right panel) that for smaller frequencies the spectral luminosity is smaller, and for larger frequencies the spectral luminosity is larger with respect to the Schwarzschild case. The results are consistent with the ones obtained in \cite{2020MNRAS.496.1115B, 2022ApJ...925..210K}, where we investigated the cases of isotropic and anisotropic pressures. This suggests that the behavior of the spectral luminosity is affected by the presence of DM in the surroundings of the black hole and the specifics of the matter model do not alter substantially the results.

In fact, one can always construct plots similar to Figs.~\ref{fig:omega_L}-\ref{fig:diff_specLum} using different DM density profile like, for instance, the Burkert profile. However no substantial difference will appear. In order to better quantify the impact of tangential pressure we present in Tabs.~\ref{tab:table1}-\ref{tab:table2} some numerical values for DM envelope using both exponential sphere and Burkert density profiles and for completeness we compare with the case where the same density profile is used for a DM envelope with isotropic pressures. 
\begin{table*}[ht]
\begin{center}
\caption{Physical parameters of test particles and DM with the exponential sphere density profile. The first column shows the values of central density given by $\rho_0=\rho_0^*\times10^{-5}\rm{AU^{-2}}$, the second column is the value of the pressure at $r=r_b$ in the presence of isotropic pressures, the third column is the radius of innermost stable circular orbits $r_{ISCO}$ in the presence  of isotropic pressures, the forth column is  $r_{ISCO}$ in the presence of tangential pressures only, the  fifth column is the surface radius of the DM envelope in the presence of isotropic pressures, which we also adopt for the tangential pressure for comparison, the sixth column is the DM mass in units of black hole mass.} 
\vspace{3 mm}
\label{tab:table1}
\begin{tabular}{|c|c|c|c|c|c|c|c|c|c|}
%\hline
\hline
\hline
 $\rho_{0}^*$ & $P_{b}$, $(10^{-8} \rm{AU^{-2}})$ & $r_{ISCO}$, (AU) &  $r_{ISCO}$, (AU)  &  $r_{s}$,(AU) &  $M_{DM}(r_{s})$, \\
             &  with isotropic pressures  &  with isotropic pressures &  with tangential pressure &   &$(10^{-2} M_{BH})$\\
\hline
\hline  $0.5$ & $2.082$ & $29.213$ & $29.141$  & $221.261$ & $1.249$ \\
\hline  $1.0$ & $4.187$ & $28.751$ & $28.679$  & $233.459$ & $2.498$ \\
\hline  $1.5$ & $6.313$ & $28.365$ & $28.218$  & $243.154$ & $3.747$\\
\hline  $2.0$ & $8.462$ & $27.954$ & $27.764$  & $253.001$ & $4.996$ \\
 \hline
 \hline
  \end{tabular}
    \end{center}
\end{table*}
%

%%%%%%%%%%%%%%%%%%%%%%%%%%%%%%%%%%%%%%%%%%%%%%%%%%%%%%%%%%%%%%%%%%%%%%%%%%%%%
%
\begin{table*}[ht]
\begin{center}
\caption{Physical parameters of test particles and DM analogous to Table~\ref{tab:table1} but with the Burkert density profile.}
\vspace{3 mm}
\label{tab:table2}
\begin{tabular}{|c|c|c|c|c|c|c|c|c|c|}
%\hline
\hline
\hline
 $\rho_{0}^*$  & $P_{b}$, $(10^{-8} \rm{AU^{-2}})$ & $r_{ISCO}$, (AU) &  $r_{ISCO}$, (AU)  &  $r_{s}$,(AU) &  $M_{DM}(r_{s})$,  \\
             &  with isotropic pressures &  with isotropic pressures &  with tangential pressure  &   &$(10^{-2}M_{BH})$\\
\hline
\hline $0.5$ & $1.208$ & $29.403$ & $29.364$  & $221.208$ & $2.264$ \\
\hline $1.0$ & $2.425$ & $29.119$ & $29.143$  & $225.571$ & $4.577$ \\
\hline $1.5$ & $3.652$ & $28.981$ & $28.917$  & $227.278$ & $6.893$ \\
\hline $2.0$ & $4.887$ & $28.745$ & $28.693$  & $230.121$ & $9.252$ \\
 \hline
 \hline
  \end{tabular}
    \end{center}
\end{table*}

In Table~\ref{tab:table1} we used the exponential sphere density profile while in Table~\ref{tab:table2} we employed the Burkert profile for comparison. For both profiles we considered two cases: with isotropic pressures and with tangential pressure only. In the case of isotropic pressures one should solve the Tolman-Oppenheimer-Volkoff (TOV) set of equations in accordance with what was done in Ref.~\cite{2020MNRAS.496.1115B}. In the case of tangential pressure the TOV equations become Eq.~\eqref{eq:diffN} in accordance with Ref.~\cite{Florides}. Since the tangential pressure is defined via Eq.~\eqref{eq:enmomten}, \eqref{eq:enmomten2} it is obvious that the pressure here is solely given by density and mass profiles. Also, since in the presence of only tangential pressure there is no need to impose vanishing pressure at the boundary the surface radius of the DM envelope $r_s$ is a free parameter. 

In order to compare the DM content in the presence of tangential pressure with the isotropic pressures case we adopted the same value for $r_s$ for both. Hence, with the same $r_s$, both cases obviously yield the same mass as shown in the sixth column of Tables  \ref{tab:table1} and \ref{tab:table2} but slight differences may appear in the location of $r_{ISCO}$ , thus influencing the accretion disk's spectrum. However one can see that, all other things being equal, the values of $r_{ISCO}$ in the two cases are comparable. 

%%%%%%%%%%%%%%%%%%%%%%%%%%%%%%%%%%%%%%%%%%%%%%%%%%%%%%%%%%%%

\section{Conclusions}\label{sez6}

Spherical matter distributions supported solely by tangential pressures may be used to describe clusters of counter rotating particles and have been used in models for relativistic compact objects for a long time\cite{Florides,kumar1970non}. The early studies of the so called `Einstein cluster' were aimed mostly at classifying the possible kinds of motion and understanding its relation to the Newtonian case. More recent studies such as Refs.~\cite{magli1} and \cite{magli2} considered dynamical scenarios and gravitational collapse which may lead to the formation of naked singularities.
In Ref.~\cite{joshi1} the `Einstein cluster' was considered as the limiting case of a collapsing cloud and its observable properties were considered in Ref.~\cite{joshi2}. 
Those models are somewhat related to our current study since they represent a DM compact object without the presence of a black hole. The properties of such objects were further investigated in \cite{joshi3}. Other authors have considered the `Einstein cluster' as a candidate for dark matter near the galactic center, most notably \cite{Harko} and more recently \cite{Harada}.

In this paper, we considered circular motion of test particles in the field of a static BH surrounded by a DM distribution of counter rotating particles described by the `Einstein cluster' metric. We computed the accretion disk luminosity spectrum and compared the numerical results with previous works that considered DM envelopes with isotropic and anisotropic pressures. We showed that the presence of the DM envelope enhances the accretion disk's luminosity.

The results we obtained from the tangential pressure picture turned out to be similar to our previous findings where we considered isotropic pressures, such as for example  Ref.~\cite{2020MNRAS.496.1115B}. The main differences appear in the value for $r_{ISCO}$,
and the value of the total DM mass which depends on the density profile adopted and its outer boundary (which is free to choose in the case of tangential pressures).
By putting together results from this work and Ref.~\cite{2020MNRAS.496.1115B} we can conclude that the dominant factor in determining the spectral luminosity of the accretion disk is the presence of a DM envelope, regardless of the chosen model for the DM itself.

While the present work offers only a qualitative picture for a static spacetime, future efforts will take into account more accurate models for galaxy, including angular momentum and adopting known observational bounds for the various free parameters of the model. In turn this may allow to put constraints on the DM distribution near galactic centers and the properties of the central supermassive compact objects.

\begin{acknowledgments}
The authors warmly thank R. D'Agostino and R. Giamb{\`o} for fruitful discussions on the topics of this work. KB, TK, EK and OL acknowledge the Ministry of Education and Science of the Republic of Kazakhstan, Grant: IRN AP08052311. DM acknowledges support from Nazarbayev University Faculty Development Competitive Research Grant No. 11022021FD2926. 
\end{acknowledgments}

\appendix

\section{Derivation of the field equations}\label{app}

Here we show the derivation of the field equations for a static and spherically symmetric matter distribution with vanishing radial pressures, following  Ref.~\cite{Florides}.
We start with the Einstein field equations given in a form
\be
Q^\alpha_\beta=G^\alpha_\beta-kT^\alpha_\beta=0,
\ee
where $G^\alpha_\beta$ is the Einstein tensor, $T^\alpha_\beta$ is the energy-momentum tensor, $k=8\pi$, $\alpha, \beta$=0, 1, 2, 3, with the same geometric units $G=c=1$ adopted throughout  the paper. 
To describe the `Einstein cluster' one can take the following components of the energy-momentum tensor
\be
T^0_0=\rho(r),\quad T^1_1=0,\quad T^2_2=T^3_3=-P_\theta(r)
\ee
where the radial pressure is assumed to be zero. Then the tangential pressure $P_\theta$ must be found from the field equations with the line element given by Eq.~\eqref{eq:le}. We get
\bea
Q^0_0&=&\frac{1}{r^2}-\frac{e^{-\Lambda}}{r^2}(1-r\Lambda')-kT^0_0=0,\label{eq:T00}\\
Q^1_1&=&\frac{1}{r^2}-\frac{e^{-\Lambda}}{r^2}(1+rN')-kT^1_1=0, \label{eq:T11}\\
Q^A_B&=&\frac{e^{-\Lambda}}{2}\left\{\left(\frac{1}{r}+\frac{N'}{2}\right)(\Lambda'-N')-N''\right\}\delta^A_B\qquad \label{eq:TAB}\\\nonumber
&&-kT^A_B=0, 
\eea
where $A$ and $B$ refer to components  2 and 3, while prime denotes derivation with respect to $r$.
Obviously, Eq.~\eqref{eq:T00} integrates to give Eq.~\eqref{eq:Lambda} while
Eq.~\eqref{eq:T11} yields Eq.~\eqref{eq:diffN}.

Now the expressions for $\Lambda'$ and $N'$ from Eqs.~\eqref{eq:T00} and \eqref{eq:T11} must be plugged in Eq.~\eqref{eq:TAB} to determine $T^A_B$. As a result, we get the tangential pressure as a function of the density profile as
\be
T_{2}^{2}=T_{3}^{3}=-P_\theta=-\frac{\rho(r)M(r)}{2\left(r-2M(r)\right)}.
\ee
It should be mentioned, that in the presence of isotropic pressure one obtains the usual set of Tolman-Oppenheimer-Volkoff equations.

\end{document}